\documentclass[10pt,journal,a4paper]{IEEEtran}

\usepackage{soul}
\usepackage{url}
\usepackage{subfigure}
\usepackage{graphicx}
\usepackage{multirow}
\usepackage{array}
\usepackage{booktabs}
\usepackage{amssymb}
\usepackage[usenames]{color}
\usepackage[usenames,dvipsnames,table]{xcolor}
\usepackage{comment}
\usepackage{mathptmx}
\usepackage{latexsym}
\usepackage{amssymb}
\usepackage{wrapfig}
\usepackage{times}
\usepackage{epsfig}
\usepackage{subfigure}
\usepackage{ifpdf}
\usepackage{lipsum}

\usepackage{hyperref}
\hypersetup{
  colorlinks=true,
  linkcolor=black,
  urlcolor=blue!70!black,
  citecolor=blue!70!black,
}
\usepackage{siunitx}
\usepackage[normalem]{ulem}
\usepackage{color}

\usepackage{glossaries}
\newacronym{api}{API}{Application Programming Interface}
\newacronym{cps}{CPS}{Cyber-Physical Systems}
\newacronym{dos}{DoS}{Denial of Service}
\newacronym{scada}{SCADA}{Supervisory Control and Data Acquisition}
\newacronym{ict}{ICT}{Information and Communications Technology}
\newacronym{lti}{LTI}{Linear Time Invariant}
\newacronym{lqg}{LQG}{Linear Quadratic Gaussian}
\newacronym{lqr}{LQR}{Linear Quadratic Regulator}
\newacronym{ml}{ML}{Machine Learning}
\newacronym{ncs}{NCS}{Networked-Control System}
\newacronym{sdn}{SDN}{Software Defined Networking}

\DeclareRobustCommand{\mybox}[2][gray!20]{%
\noindent
\fcolorbox{gray!25}{gray!7}{
\parbox{1.9\columnwidth}{
  #2
}}
\medskip
}

\title{Next Generation Resilient Cyber-Physical Systems}

\author{Michel Barbeau,~\IEEEmembership{Member~IEEE},
Georg Carle,~\IEEEmembership{Member~IEEE},\\
Joaquin  Garcia-Alfaro,~\IEEEmembership{Senior~Member,~IEEE},\\ and
Vicen\c{c} Torra,~\IEEEmembership{Fellow~Member,~IEEE}

\thanks{M. Barbeau, Carleton University, School of Computer Science,
Canada. E-mail: barbeau@scs.carleton.ca}

\thanks{G. Carle, Technical University of Munich, Department of Informatics,
Chair of Network Architectures and Services, Germany. E-mail:
carle@net.in.tum.de}

\thanks{J. Garcia-Alfaro, Institut Polytechnique de Paris,
CNRS UMR 5157 SAMOVAR, T\'el\'ecom SudParis, France. E-mail:
jgalfaro@ieee.org}

\thanks{V. Torra, National University of Ireland, Maynooth,  Hamilton
Institute, Ireland. E-mail: vtorra@ieee.org}}

\begin{document}

\thispagestyle{plain}
\pagestyle{plain}

\maketitle

\noindent \begin{abstract} A distributed \gls*{cps} consists of
  tightly integrated computing, communication and control
  technologies. Recent \gls*{cps} hacking incidents had significant
  consequences. In such a context, reinforcing their resilience,
  referring to their capacity to recover from disruptions, is a key
  challenge. Concretely, it involves the integration of mechanisms to
  regulate safety, security and recovery from adverse events,
  including plans deployed before, during and after incidents occur.
  We envision a paradigm change where an increase of adversarial
  resources does not translate anymore into higher likelihood of
  disruptions. Consistently with current system design practices in
  other areas, employing high safety technologies and protocols, we
  outline a vision for next generation \gls*{cps} addressing the
  resilience challenge leveraging ideas such machine learning and 
  fuzzy systems.
\end{abstract}

\section{Introduction}
\label{sec:intro}

\noindent \IEEEPARstart{C}{y}ber-Physical Systems (CPS) integrate
computation, communication and physical processes~\cite{kim2012cyber}.
The design of a \gls*{cps} involves several fields including computer
science, control theory, automation, networking and distributed
systems. Skills from these domains are put together to ensure that a
myriad of computing resources and physical elements get orchestrated
via networking technologies. In addition, \gls*{cps} integrate
facilities for human-computer interaction. Examples of \gls*{cps}
interacting with humans include \emph{industrial control systems}
(e.g., workers operating industrial machines) and \emph{smart cities}
(involving thousands of nodes and citizens).

\gls*{cps} are omnipresent in our everyday life. Hacking and failures
of such systems have impact on critical services with potentially
significant and lasting consequences. Reinforcing their resilience is
a key challenge. Resilience refers to the capacity of a system to
recover from disruptions. It can be seen as the mechanisms present in
a system to regulate its safety and security and to recover from
adverse events. Resilience includes actions and plans that are
deployed before, during and after adverse events take place.
Resilience is a historical term used as a descriptor in complex
fields, from psychology and medicine to civil and military
engineering. In cybersecurity, it relates to the idea of how a complex
system bounces back from a disruption, as well as all the possible
post-disruption strategies followed after the events are recognized.

\begin{table*}[!hptb]
\centering
\centering
\caption{Representative cyber-physical attacks reported in the media.}
\label{tab:incidents}
\mybox{
\begin{tabular}{ p{7.95cm} | p{7.95cm}}
\textbf{1. Sabotage of critical facilities}, such as a
\href{http://j.mp/2otvCn1}{German steel mill} in 2015, hospitals,
media and financial services in \href{http://j.mp/2WwXMO0}{France and
  the UK} in 2017 and 2018. The problem is spanning several countries
from the European Union, the US, \href{http://j.mp/wannacry2018}{and
  beyond}. &

\textbf{1. Human adversarial actions in this scenario}, include USB
injection of corrupted software binaries, drive-by-download malware
installation, spear phishing-based design of websites, and traditional
social engineering manipulation of critical infrastructure
employees.\\

\medskip

\textbf{2. Remote control of navigation systems}, including successful
hacking of \href{http://j.mp/2yW61Dr}{autonomous cars} and
\href{http://j.mp/2B1RtrR}{avionic systems}. Studies and general
concern started with a malware that infected over sixty thousand
computers of a \href{http://j.mp/2jaM6uM}{Iranian nuclear facility},
and destroyed more than one thousand \textbf{nuclear centrifuges}.
This delayed the Iran's \textbf{atomic program} by at least two years.
&

\textbf{\newline \newline 2. Human adversarial actions} include the
use of infection vectors (e.g., USB drives), corrupted updates and
patches, radio frequency jamming, radio frequency spoofing, and
software binary manipulations.\\

\medskip

\textbf{3. Disruptions of large-scale industries} have been appointed by
the Federal Office for Information Security of Germany as a serious
concern to European factory and industrial markets. Similar threats
affect~\href{http://bit.ly/2LMqR3H}{drones and smart cities}, as well.&

\textbf{\newline 3. Human adversarial actions} include the use of GNSS
(Global Navigation Satellite Systems) attacks, such as jamming and
spoofing of signals, and hijacking of communications to downgrade
communications to insecure modes (e.g., downgrading from encrypted to
plain-text communications).\\

\end{tabular}
}
\end{table*}

\subsection{Related Work}
\label{sec:relatedwork}

The state of the art in \gls*{cps} security has recently been reviewed
by Girlado et al.~\cite{Giraldo2017} and Humayed et
al.~\cite{Humayed2017}. According to Giraldo et al., past research
works have put much emphasis on the problem of preventing perpetration
of attacks on \gls*{cps}, for instance, leveraging cryptographic
techniques and building intrusion detection systems. They emphasize
the need for more works on techniques for mitigating the consequences
of attacks. After they have been detected, the problem of responding
to attacks seems to have received little attention. Humayed et al. did
a good job at identifying representative \gls*{cps} and reviewing
security issues specific to them. The categories are industrial
control systems, medical devices, smart cars and smart grid systems.
For every representative \gls*{cps}, specific threats,
vulnerabilities, attacks and controls are examined.

Although the term \gls*{cps} emerged recently, it builds upon very 
well-established
research fields, i.e., embedded computing, control theory and
human-computer interaction. For instance, a \gls*{cps} can be easily
modelled as a \gls*{ncs}~\cite{ncsCarle18}. The major difference is
that the controller is coupled with the actuators and sensors through
a communication network (e.g, an Ethernet-like network). The use of
this communication network to connect the components provides
flexibility and low implementation costs~\cite{gupta2010networked}.
\gls*{ncs} classical theoretical problems include (1)~stabilization of
system processes given delays and packet losses due to the network
elements~\cite{wang2008networked, zhang2001stability}; (2)~data rate
limiting techniques (e.g., control to systems
traffic)~\cite{hespanha2007survey} and (3)~energy efficiency for
wireless \gls*{ncs}~\cite{antsaklis2007special,tiberi2013energy}. It
is only until recently that the \gls*{ncs} communities started working
on cybersecurity issues of \gls*{cps}~\cite{kim2012cyber,wu2016survey}.
Obviously, the use of a communication network to transport control and
observations, i.e., signals to actuators and from sensors, paves the
way to important security vulnerabilities~\cite{smith2015covert}. A
\gls*{ncs} can be attacked and needs to be protected.

Attacks exploiting \gls*{ncs} vulnerabilities can be characterized
according to three main aspects~\cite{teixeira2012attack}:
(a)~adversary's \emph{a priori} knowledge about the system and its
protective measures, (b)~class of disrupted resources (e.g.,
denial-of-service attacks targeting elements that are crucial to
operation) and (c)~analysis of control signals during perpetration of
an attack (e.g., sensor outputs), that may be used to carry out more
sophisticated attacks (e.g., attacks targeting the integrity or
availability of the system). The \textit{knowledge of adversaries} in
terms of, e.g., system dynamics, feedback predictability and system
countermeasures, can be used to perpetrate attacks with severe
security and safety implications, when they target the operations of,
e.g., industrial systems and national infrastructures. They can lead
to \textit{catastrophic consequences to businesses, governments and
  society at large}. A growing number of attacks on cyber-physical
infrastructures are reported in the world, targeting vital activities
(e.g., water, energy and transportation) for intelligence or sabotage
purposes. Some representative incidents are outlined in
Table~\ref{tab:incidents}.

A careful review of incidents as those in Table~\ref{tab:incidents}
reveals that they all have a common element~\cite{humayed2017cyber,
  pasqualetti2015control}: \textit{human adversarial actions} forging
system feedback measurements for disruption purposes~\cite{gasser18}.
The underlying issue, hereinafter called the \textbf{\emph{feedback
    truthfulness problem}}, refers to \textit{intentional situations
  perpetrated by human adversaries, forging physical observations in a
  stealthy manner}~\cite{arghandeh2016, iturbe2016feasibility}. They
are cyber-physical attacks generating
anomalies~\cite{hartung2015prescription, smith2015covert}. However,
even if detected, the attacks appear as unintentional errors. Hence,
they are leading to wrong resilience plans. \emph{How to distinguish
  an intentional attack from an unintentional fault?} This a challenge
because symptoms may almost be the same, but reactions should be
different. Indeed, the correct response to a fault is a repair action
that restores the state of the system. In case of an intentional
attack, physical resources may not be faulty at all, but the adversary
makes them appear faulty. A repair action will not help.

It is crucial to address the aforementioned challenge in a provable
manner in order to prioritize \textit{appropriate responses} and
rapidly recover control to assure \textit{cyber-physical
  resilience}~\cite{garcia2015CEE, humayed2017cyber,
  kieseberg2018security}. That is, to assure the persistence of the
system when facing changes, either accidental or
intentional~\cite{uday2015designing}. In terms of \gls*{cps} design,
cyber-physical resilience shall also deal with the management of
operational functionality that is crucial for a system, and that
cannot be stopped. In other words, system functionality that shall be
properly accomplished. Regarding the incidents mentioned in
Table~\ref{tab:incidents}, the cooling service of reactor in a nuclear
plant, or the safety controls of an autonomous navigation system, are
proper examples of critical functionalities. Other system
functionalities may be seen as less important; and even be temporarily
stopped or partially completed. Such type of functions can be seen as
secondary. A \textit{printing service for employees} in a nuclear
plant scenario is a proper example of a \textit{secondary function}
that \textit{one might accept to sacrifice}, under graceful
degradation.

When addressing resilience, two crucial elements to take into
consideration are \textit{the severity of the actions disrupting the
  functionalities of a system} and \textit{properly distinguishing
  accidental failures from intentional attacks}~\cite{SPpanel2014,
  iturbe2016feasibility, krotofil2015process, rauter2018integrating,
  yu2017guest}. The objective is to use the proper \textit{security
  stacks} and deploy \textit{resilience plans}, including responses
that mitigate the impact of undesirable actions~\cite{garcia2015CEE}.
This includes the use of \textit{proactive}, often short-term,
\textit{tactical policies to handle failures}; and \textit{reactive},
usually long-term, \textit{strategies for
  attacks}~\cite{huang2017integration, motzek2017selection,
  qi2015interaction}. Security stacks in both areas can include
redundancy (e.g., use of additional system replicas),
compartamentalization, segmentation, and activation of upgraded modes
of protection (e.g., use of cryptography to enable secure handshakes,
message signatures, and encryption\cite{Giraldo2017, Humayed2017}).
The inclusion of resilience plans shall always keep critical processes
in a normal operating mode, while the system is confronted with
incidents. The challenge of satisfying those requirements on automated
\gls*{cps} designs stresses the importance of determining the root
nature of incidents, to drive the appropriate models (e.g., in terms
of remediation) that the system must select and enforce in the end.

One may also consider that both accidental failures and intentional
attacks can be formally represented as anomalies in the measured data.
Recent studies by Iturbe-Urretxa et al.~\cite{iturbe2016feasibility,
  iturbe2017data} discuss on the feasibility of distinguishing between
process disturbances and intrusions in process control systems using
multivariate statistical process control. More specifically, the
authors define a statistical analysis process for the definition of
normal traffic, reporting anomalies (i.e., deviations from the
expected profile of a trustworthy entity) as adversarial activities.
Authors show the way of dealing with the complexity of data management
as a result of monitoring processes collecting and transforming
anomalous events in industrial control systems mathematically modeled
as \gls*{ncs}. The contributions of the study range from visual
analytics to detection and correlation of anomalous events based on
statistical management of large datasets.

\medskip

More relevant related works and reference material are cited
throughout the remaining sections of the paper.
Section~\ref{sec:resilience} provides a more thorough introduction to
the concept of resilience and the use of security stacks to enable
cyber-physical protection. Section~\ref{sec:survey} argues
the necessity of a paradigm change and discusses our vision of how
next generation resilient \gls*{cps} will be addressing such a change.
Section~\ref{sec:conclusion} closes the paper.

\section{Resilience and Security Stacks}
\label{sec:resilience}

Resilience is a term with centuries of use~\cite{linkov2019}. It
encompasses multidisciplinary fields, from psychology and medicine to
civil and military engineering. Current application of the term under
the scope of cybersecurity is centered upon the idea of bouncing back
from failures, while defending forward from attacks. It emphasizes the
capacity of a system to recover from disruptions, and is often seen as
the underlying technique by which a system regulates its safety and
security mechanisms, to recover from adverse events. Resilience
includes actions and plans that must be conducted before, during, and
after events take place. Resilience is a historical term used as a
descriptor in complex fields, from psychology and medicine to civil
and military engineering. The modern application of resilience relates
to the idea of how a complex system bounces back from a disruption, as
well as all the possible post-disruption strategies that may come
after the events are identified.

Under the scope of complex systems theory, the concept of resilience
may be confused with other traditional concepts such as robustness,
fault tolerance and sustainability. However, there exist fundamental
differences between such terms. For instance, while robustness stands
for the ability to withstand or overcome adverse conditions (e.g.,
faults and attacks), resilience refers more to the capacity for a
system to maintain functionality despite the occurrence of some
internal or external disruptions, e.g., adversarial
breach~\cite{dhsgov2010}. Similarly, fault tolerance refers more to
the maintenance of crucial services within a given time-period under
the presence of failures and sustainability to similar metaphors in
disciplines like environmental and socio-ecological
processes~\cite{ahern2011fail, gunderson2000}.

Laprie~\cite{laprie2008dependability} settled some key
definitions when comparing resilience to dependability and fault
tolerance. In his work, Laprie related the resilience and
dependability terms as follows: \textit{Resilience is the persistence
  of dependability when facing changes}. More recently, the relation
between resilience and performance targets have been described by
Meyer~\cite{meyer2009defining} as follows: \textit{Resilience is
  the persistence of Performability when facing changes}. This can
be accomplished by \textit{graceful degradation}, i.e., by
prioritizing some services over non-essential ones, for as long
as possible~\cite{gonzalez1997}.

The concept of resilience spans across several other disciplines. For
instance, when talking about resilience in terms of network theory,
resilience refers to the persistence of service delivery when the
network faces changes~\cite{yazdani2011resilience}. In terms of
quality of service, resilience
relates to the degree of stability of the services provided by the
system~\cite{autenrieth2002using}. From a control-theoretic
standpoint, resilience refers to the
ability to reduce the magnitude and duration of deviations from
optimal performance levels~\cite{pasqualetti2015control}. Finally,
resilience is also seen in
disciplines such as medicine and psychology, as the ability to recover
from a crucial trauma or crisis~\cite{yates2004fostering}. The common
element seen in all the
aforementioned definitions relates to adaptation to confront change
and significant adversities.

When we move to the specific context of cybersecurity, resilience
means accepting that the system is vulnerable to attacks, in addition
to faults and failures~\cite{huang2017integration,arghandeh2016}. It
means to accept that there will be breach
of security (e.g., by a collusion between insiders and outsiders,
attacking and disrupting the system). Handling resilience in the
cybersecurity context means holding an adversarial mindset and getting
ready to lose some assets~\cite{gonzalez1997}. This does not mean
sacrificing the system,
but deciding which parts of the system we can lose (accepting that we
must lose some control over the system) while prioritizing those
assets we must give up to assure that the system will remain
functional during the disruptions.

To improve resilience from the
cybersecurity standpoint relies on enforcing a traditional security
stack, in terms of identifying the system weaknesses (e.g., in their
software and infrastructure themselves) that could potentially be
controlled by a skilled adversary with the purpose of disrupting the
system. Management in terms of identifying vulnerabilities must be
followed as well by assessment of incidents, service continuity and,
in general, any risks affecting the system. These aforementioned
management perspectives must be driven by resilience thinking in the
form of \textit{bouncing back} (or defending back) from disruptive or
adverse events. In other words, attacks against the availability of a
given service, as well as any incident
leading to security breaches must be quickly solved (e.g., incidents
must properly be absorbed). 

\section{Moving Forward}
\label{sec:survey}

\noindent In the previous section, we argued that modern \gls*{cps}
must change today's adversarial paradigm where an increase in the
resources of the adversaries always translates into higher likelihood
of disruption. In this section, we survey some promising techniques
that could potentially help the dynamics of the game. All computer
based systems can take advantage of these techniques to improve their
safety and security. In the context of \gls*{cps}, we discuss how each
of these techniques can improve security.

\medskip

\noindent \textbf{Machine Learning ---} Artificial Intelligence (AI)
by means of the subfields of \gls*{ml} and search provides a large set
of techniques appropriate for resilient cyber-physical systems. There
are three main \gls*{ml} paradigms, namely, supervised, unsupervised
and reinforcement. In supervised machine learning, there are old and
new data points. Old data points are labelled. A label represents a
classification of data points. Comparing their similarity with old
data points, supervised machine learning assigns classes or labels to
new data points. With unsupervised \gls*{ml}, the data points are
unlabelled (i.e., learning is about extracting information from data).
Data points are grouped together into classes according to similarity.
The classes need to be labelled by a human expert. In contrast,
reinforcement learning rewards or penalizes the learner following the
validity of inferred classifications, i.e., there is no need for
labelled data. Learning is inferred from the successes and failures.

Supervised and reinforcement \gls*{ml} is used for system
identification and model fitting. Different alternative learning
methods exist, based on different considerations on the type of the model
(e.g., rule-based, support-vector machines, deep learning models) and
its properties (e.g., explainable models/decisions, efficiency).

Resilience plans build upon rational responses. Their performance
often requires rapid completion of search tasks. Their efficiency can
be greatly improved when the search are informed, i.e., when it
applies heuristics. The AI subfield of \emph{search} provides us with
algorithms and methods for complex decision making problems. For
example, systems based on Monte Carlo tree search have been proven
successful in difficult games (e.g.,
\href{https://en.wikipedia.org/wiki/AlphaGo}{AlphaGo} and
\href{https://en.wikipedia.org/wiki/AlphaZero}{AlphaZero}). Connection
between Monte Carlo tree search and reinforcement learning exists in
the AI literature~\cite{silver2009reinforcement,vodopivec2017monte}.

How does \gls*{cps} security can take advantage of \gls*{ml}? The
start of an answer can be found in a book authored by Chio and
Freeman~\cite{chio2018machine}. It is worth mentioning that the
applicability of \gls*{ml} to computer security has been demonstrated
in the past. The most successful story is the use of the approach to
control spam emails. Metadata, source reputation, user feedback and
pattern recognition have combined to filter out junk emails.
Furthermore, there is an evolution ability. The filter gets better
over time. \gls*{ml} is about data and, together with clever
algorithms, building experience such that next time the system does
better. This way of thinking is relevant to \gls*{cps} security
because its defense can learn from attacks and make the
countermeasures evolve. Focussing on \gls*{cps}-specific threats, as
an example pattern recognition can be used to extract in data the
characteristics of attacks and prevent them in the future. Because of
its ability to generalize, \gls*{ml} can deal with adversaries hiding
by varying the exact form taken by their attack. Note that
perpetrators can also adopt the \gls*{ml} paradigm to learn defense
strategies and evolve attack methods. The full potential of \gls*{ml}
for \gls*{cps} security has not been fully explored. The way is open
for the application of \gls*{ml} in several scenarios.

\medskip

\noindent \textbf{Fuzzy Decisional Systems ---} Fuzzy sets can be
used to model imprecision and vagueness. A concept is said
imprecise when several values satisfy it (e.g., the temperature is
below zero). A concept is vague when it represents partial truth. For
example, the fact that a temperature is {\em near zero} can be a
matter of degree and there is no value under which temperatures are
near zero and over which it is completely false that the temperatures
are near zero. Fuzzy systems are typically rule based systems in which
concepts are represented by means of fuzzy sets. This permits that in
particular situations, terms are partially fulfilled and, as a
consequence, rules are partially fired.

Fuzzy sets have been proven to be effective in modelling safety and
control. Fuzzy control being one of the most successful application
areas of fuzzy sets. In these applications, a control system is
defined by a set of fuzzy rules that will be fired all at once. The
set of consequents of all rules are then combined taking into account
the partial fulfillment of each rule. Combination results into a fuzzy
set that needs to be defuzzified to result into an actual value. When
the number of variables in a system become large, the construction of
fuzzy sets systems need to deal with the course of dimensionality, as
the number of rules are typically exponential on the number of
variables. Hierarchical fuzzy systems have been developed to deal with
this problem. Adaptive systems exist that modify the rules according
to changes in the environment. Fuzzy rules can be learned from data
and, thus, used for adversarial identification. Fuzzy rule based
systems can be efficiently deployed in real time systems. This is so
because rules can be fired in parallel and inference can be also
implemented in an efficient way. Fuzzy systems can also be used to
model high-level decision making processes, as e.g. to reason about
identification of adversarial actions, and the remediation to be
taken. These decisions need to take into account high doses of
uncertainty.

The potential of fuzzy decision making in computer system security has
been demonstrated. Its ability to deal with uncertainty is
particularly useful for risk
assessment~\cite{ALALI2018323,Erdogan2020}. Normal operation
conditions of a CPS are a vague concept and detection of abnormal
conditions and unstable states can be modelled and inferred using
fuzzy systems where partial truth is accommodated. Fuzzy systems are
also useful for adaptive control environments, in which the underlying
models (e.g., system dynamics and related parameters) vary frequently,
due to the high degree of uncertainty in the system. All this needs to
be combined with probabilistic approaches as attacks are either
present or absent and thus better represented with probabilistic
approaches (i.e., attacks being related to intentionality while
failures and faults associated to uncertainty).

\section{Conclusion}
\label{sec:conclusion}

A \gls*{cps} is a physical process observed and controlled through a
computer network. Signals to actuators and feedback from sensors are
exchanged with a controller using a network. The advantages of such an
architecture are flexibility and relatively low deployment cost. A
\gls*{cps} will always be prone to failures and malicious attacks. The
networking aspect of \gls*{cps} opens the door to cyber-physical
attacks. Analysis of past incidents highlights the advanced knowledge
degree of the adversaries perpetrating the attacks. Adversaries are
smart and they can learn. Their sophistication is such that they can
fool the controllers forging false feedback. Hence, a fundamental
\gls*{cps} security problem is the feedback truthfulness.

The first burning question is the need to distinguish an unintentional
failure from a malicious attack. The signs resulting for these
undesirable situations may be the same, but the responses should be
different. A fault can be repaired. Against, an attack a \gls*{cps}
has to defend itself. Acknowledging that the operation of a \gls*{cps}
may be disrupted by a malicious attack, the second burning question is
building a \gls*{cps} with resilience. That is, it must be able to
recognize the presence of an attack, recover and maintain operation.
Several stories of attacks and disruption told in the media (see
Table~\ref{tab:incidents}) are evidence of the relevance of the
problem and the increasing risks of major catastrophes in sectors such
as industry, manufacturing, transport or power generation. Currently,
\gls*{cps} are in principle secure by design, in the sense that they
implement state of the art cryptography and protection techniques. In
the future, they need to be resilient by construction. We introduced
the defense learning paradigm where knowledge is built about
adversaries, their techniques are identified, weaknesses are
discovered, actions are anticipated and transformed into regular
actions.

We have presented our vision on how next generation resilient
\gls*{cps} will be. The same way that nobody can think about current
\gls*{cps} without perfect safety to argue resilience; we have claimed
that in some years, nobody would think about a \gls*{cps} without
perfect cyber-physical protection, in which the adversarial paradigm
would have to change and make sure that an increase in the adversarial
resources does not translate into higher likelihood of \gls*{cps}
disruption. We have also listed some promising techniques promoted by
artificial intelligence (AI) and machine learning (\gls*{ml})
communities, that may materialize the new security stack addressing
security beyond breach. We believe AI/ML, heuristic search and 
fuzzy decisional systems will play roles in the design of \gls*{cps}
resilience.

The essence of the war between adversaries and defenders is knowledge.
On the one hand, supervised and reinforcement learning can be used by
an adversary for the purpose of system identification, an enabler for
covert attacks. On the other hand, the design of resilience plans can
leverage AI heuristic search to speedup decision taking during the
execution of a resilience plan. The adaptive control that resilience
requires may be obtained using the fuzzy decisional approach. Quantum
techniques can eventually perform searches with time complexity that
is data size independent.

\bibliographystyle{plain}
\bibliography{main}

\end{document}